\documentclass[preprint,showpacs,preprintnumbers,amsmath,amssymb
               superscriptaddress]{revtex4}

\usepackage{graphicx}% Include figure files
\usepackage{dcolumn}% Align table columns on decimal point
\usepackage{bm}% bold math

\begin{document}

\preprint{\today, Version 01}

% -------------------------------------
\newcommand{\bb}{{\bf b}}
\newcommand{\beq}{\begin{equation}}
\newcommand{\eeq}{\end{equation}}
\newcommand{\bB}{{\mathbf{B}}}
\newcommand{\bE}{{\bf E}}
\newcommand{\bH}{{\bf H}}
\newcommand{\bD}{{\bf D}}
\newcommand{\bM}{{\bf M}}
\newcommand{\bN}{{\bf N}}
\newcommand{\bL}{{\bf L}}
\newcommand{\br}{{\bf r}}
\newcommand{\bR}{{\bf R}}
\newcommand{\bc}{{\bf c}}
\newcommand{\bq}{{\bf q}}
\newcommand{\bu}{{\bf u}}
\newcommand{\bV}{{\bf V}}
\newcommand{\murb}{{\bar{\mu}_r}}
\newcommand{\etheta}{{\bf e}_{\theta}}
\newcommand{\ephi}{{\bf e}_{\phi}}
\newcommand{\er}{{\bf e}_{r}}
\newcommand{\ex}{{\bf e}_{x}}
\newcommand{\ey}{{\bf e}_{y}}
\newcommand{\ez}{{\bf e}_{z}}
\newcommand{\ep}{{\bf e}_{+}}
\newcommand{\emm}{{\bf e}_{-}}
\newcommand{\eo}{{\bf e}_{0}}
\newcommand{\uvmn}{\,^{uv}_{mn}}
\newcommand{\mnuv}{\,^{mn}_{uv}}
\newcommand{\mnpq}{\,^{mn}_{pq}}
\newcommand{\dsum}{\displaystyle\sum}
\newcommand{\wtc}{\widetilde{c}}
\newcommand{\wtd}{\widetilde{d}}
\newcommand{\barc}{\bar{c}}
\newcommand{\bard}{\bar{d}}
\newcommand{\bark}{\bar{k}}
\newcommand{\tg}{\widetilde{g}}
\newcommand{\te}{\widetilde{e}}
\newcommand{\tf}{\widetilde{f}}
\newcommand{\barg}{\bar{g}}
\newcommand{\bare}{\bar{e}}
\newcommand{\barE}{\bar{E}}
\newcommand{\barf}{\bar{f}}
\newcommand{\breg}{\breve{g}}
\newcommand{\bree}{\breve{e}}
\newcommand{\bref}{\breve{f}}
\newcommand{\cG}{{\cal G}}
\newcommand{\cE}{{\cal E}}
\newcommand{\cF}{{\cal F}}
% -------------------------------------

\title{Grain boundary roughening transitions}

\author{S T Chui}
\affiliation{Bartol Research Institute and Dept. of Physics and Astronomy, 
University of Delaware, Newark, Delaware 19716}
% -------------------------------------------

\date{\today}

\begin{abstract}
We consider the roughening of small angle grain boundaries
consisting of arrays of dislocations and found two transitions, 
corresponding to fluctuations of the dislocations along and perpendicular
to the boundaries. The latter contributes to a large scale fluctuation
of the orientation of the crystal but the former does not.
The transition temperatures of these transitions are
very different, with the latter occuring at a much higher temperature. 
Order of magnitude estimates of these temperatures are consistent with
recent experimental results from elasticity and X-ray  measurements
in solid $^4$He.
\end{abstract}

\pacs{
67.80.-s
}

\maketitle
Since the discovery of an increase in torsional oscillator frequency in solid 
$^4$He at around 200 mK\cite{chan}, there have been renewed interests in its 
low temperature physical properties. Much recent focus is on the role
played by defects in this system.\cite{x,jb,balibar,reppy,dis,toner,lagb}
For example, it is suggested that large angle grain boundaries can
exhibit superfluid behavior\cite{lagb,balibar}.
X-ray measurements\cite{x} found a change
in the orientational fluctuation of the crystallites at around 1.75 K.
Recently Day and Beamish\cite{jb}
found a change in the shear modulus with the same temperature dependence
as that for the decoupling in torsional oscillators. They 
ascribe this to a change in the mobility of dislocations. 

There usually is a {\bf finite} density of dislocations and it is important
to consider the long range elastic interaction between them.
The simplest arrangement of a collection of dislocations comes
from the small angle grain boundary (GB).  
These boundaries are pinned by the
Peierls potential even in the absence of additional impurities. 
There has been much interest in the study of the pinning of an elastic
two dimensional interface. Above the roughening transition temperature $T_R$
the free energy to create a step becomes zero and the interface is depinned.
This roughening transition temperature is a function of the strength of the
pinning potential. Even as the strength of the pinning potential approaches
zero, $T_R$ remains {\bf finite}.

In this paper we examine if a GB can roughen and found that
for a "electrically neutral" system, there are {\bf two} 
roughening transitions, corresponding to the motion of the dislocations
parallel and perpendicular to the boundary. 
The latter contributes to a large scale fluctuation
of the orientation of the crystal as is observed in the X-ray
experiments\cite{x}.  The other transition occurs at a much 
lower temperature, does not contribute to the large scale angular
fluctuations but, because of the change in mobility of the dislocation,
can cause a change in the elastic coefficients, similar to that 
observed experimentally\cite{jb}.
Order of magnitude estimates of these temperatures are consistent with
experimental results from elasticity\cite{jb} and X-ray  measurements\cite{x}
in solid $^4$He. We now describe our results in detail.

A small angle GB consists of an array of dislocations
with parallel Burger's vectors \bb.
The trajectory of a dislocation can be represented by the positions
of elements separated by lattice constants $a_z=a_0$ along it.
We describe the configuration of the GB by the positions
$\bc_j$ of the elements of the dislocations. 
The elastic interaction between two elements of the dislocations
is given by the formula 
\beq
V=\kappa /(4\pi)[ \bb\cdot\bb' /R + \bb\cdot \bR\bb'\cdot \bR/R^3]
\eeq
where $\bR$ is the distance between the elements.
For simplicity we assume an elastically isotropic system.
In terms of the Lame constants $\lambda$ and $\mu$,
$\kappa =4a_0^2\mu(\mu+\lambda)/(2\mu+\lambda)$.
In addition, for each dislocation there is a
core energy contribution $V^c=(E_c/a_0)L$ that is
proportional to its length $L=\int  dl,$ where $dl=(dc_x^2+dc_y^2+dc_z^2)^{1/2}.$
Finally there is the pinning energy which
we assume to be of a Peierl's form 
$\sum_j U_x\cos (c_{xj}/a_0)+ U_y\cos (c_{yj}/a_0).$
\begin{figure}%[htb]
%\begin{center}
\includegraphics[width=6cm]{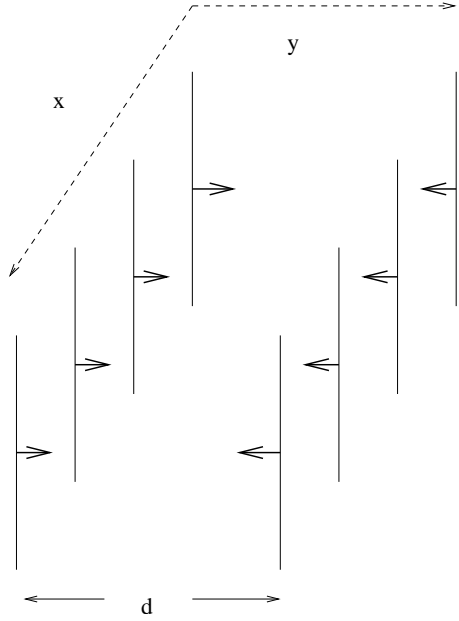}
%\end{center}
 \caption{\label{fig1} Schematic diagram of the grain boundaries.}
\end{figure}

We take the dislocation line along z; the Burger's vectors along y. 
Thus the GB is in the xz plane. This is illustrated in fig. 1.
The roughening transition depends on the form of 
the energy change for small devaitions $\delta \bc_j$
in the location of the dislocations. This deviations of the dialocations
%along the z direction 
are along the x (parallel to the GB) 
and y (perpendicular to the GB) directions only.
The fluctuation in the location of the
dislocations $\delta \br$ does not have a component along z.
The energy change determines the "elastic"
properties of the GB. We shall be interested in the Fourier transform 
$\delta \bc (q) =\sum_j \delta \bc_j \exp(-iq\cdot c_j)/\sqrt{N}.$
%$\delta \bc_j=\sum_q \delta \br_q \exp(i\bq\cdot \bc_j)/\sqrt{N}.$
The GB is in the xz plane. Hence the wave vector $\bq$
is in the xz plane only.
The distance between the dislocations $a_x=a_0/\theta$ is related
to the misorientation angle $\theta$ between the grains.
We consider the simplest "electrically neutral" system 
with zero total Burger's vectors consisting of two
grain boundaries of opposite Burger's vectors; one at y=0 and the other
one at y=d.
We focus on the lowest energy 
acoustic mode so that the displacements
on the two boundaries are the same.

A single grain boundary corresponds to two {\bf infinite} crystals of different
orientation joining together whereas a grain will be surrounded by
boundaries of opposite orientations. We find that a {\bf single} grain boundary 
does {\bf not} roughen and thus focus on the pair, which also occur physically.

The contribution to the energy change from the core energy is equal to
\beq
\delta V^c=\sum_{\bq}0.5E_c q_z^2 [|\delta c_x(\bq)|^2+|\delta c_y(\bq)|^2].
\eeq
%As an estime, take $E_c=\kappa b^2/(2\pi).$

We next look at the contribution from $V$.
From eq. (1) the change in V is given by
\beq
\delta V=\kappa /(4\pi)b^2\sum_{i,j=x,y}\delta c_{iq}\delta c_{j,-q}
(D^s_{ij}+D'_{ij})
\eeq
where the contribution  from dislocations in the same GB (self) is
\beq
D^s_{ij}=\sum_R [1-\cos(iq\cdot R)]
[\nabla_i\nabla_j 1/R +\nabla_i\nabla_j y^2/R^3]
\eeq

Similarly the contribution from the
interaction energy between dislocations on different boundaries is given by
\beq
D'_{ij}=-\sum_R [1-\cos(iq\cdot R)]
[\nabla_i\nabla_j 1/R'+\nabla_i\nabla_j U]
\eeq
$U=(y+d)^2/R'^3,$
$r'=(R^2+d^2)^{1/2}.$
$D$ is very similar to the dynamical matrix
for the two dimensional Wigner crystal which has been
considered in detail with the Ewald sum technique by 
Bonsall and Maraduddin\cite{BM}.
The two dimensional sums $D^s$, $D'$ can be evaluated in the same manner.

We find in the long wavelength limit the x mode and the y mode are not
coupled.  For the "x" mode, the energy change is 
\beq
\delta V^c+\delta V^x=(C_xq_x^2+C_zq_z^2)|\delta c_x(\bq)|^2,
\eeq
$C_x=\kappa b^2(\pi^{-1/2}+2)d/(a_xa_z),$ %=6.3 \kappa  b^2/a_z ,$
$C_z=0.5E_c-0.21 \kappa  b^2/(a_z\pi).$ Without the core energy contribution
$C_z$ is less than zero and the lattice is unstable.
This is consistent with the fact that a 2D rectangular Wigner lattice is
unstable.
Similarly for the "y" mode, the energy change is 
\beq
\delta V^c+\delta V^y=(C'_xq_x^2+C'_zq_z^2)|\delta c_y(\bq)|^2
\eeq
where
$C'_x=\kappa b^2 (\pi^{-1/2}+7/4)d/(a_xa_z),$ %=3.86 \kappa b^2/a_z ,$
$C'_z= 0.5E_c+3.6 \kappa b^2/a_z.$
The "elastic constants" $C$ for the x mode 
are much smaller than that for the y mode.
When the pinning energy
are included, the energy we get is of the same form
as that in the study of the roughening transition. 

For a single unpaied grain boundary, the elastic energy is of the {\bf first}
power in $q$. The cost of the long range fluctuation is higher and 
these fluctuations are supressed. As a result, an unpaired grain
boundary does not roughen.
We next estimate the roughening temperatures of our system.
\begin{figure}%[htb]
%\begin{center}
\includegraphics[width=8cm]{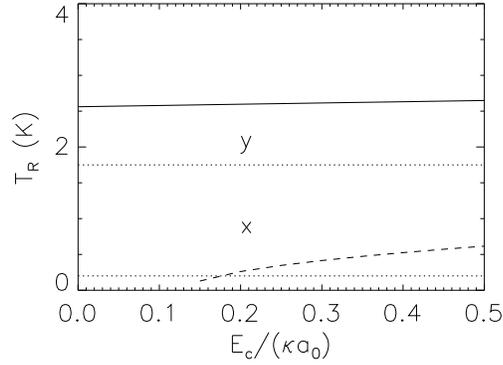}
%\end{center}
 \caption{\label{fig1} The roughening temperature in units of Kelvin
for the y mode (solid line) and the x mode (dashed line) as a function
of the core energy normalized by $\kappa a_0.$ The experimental
transition temperatures are indicated by the dotted line.}
\end{figure}

From the study of the roughening transition we find that, 
as the pinning Peierls potential approaches zero, the
roughening transition is given by
\beq
kT_c=2a_zC/a_x
\eeq 
where $C$ is the geometric mean of the elastic constants. For example,
for the x mode $C=(C_xC_y)^{1/2}.$
The crystal symmetry of $^4$He is HCP.
In the present calculation, we have approximated it by an 
elastically isotropic system.
Our goal is not to make an accurate prediction of the temperature but
to find out if the physics discussed in this paper is of relevance to the
experimental system.  We have used the following estimates
of the Lame constants\cite{elas}
$\mu=0.72 K/A^3.$
$\lambda= 1.7 K/A^3.$
%Take a density He4 of $2.2\times 10^22/cm^3 =1/a^3,$ we get
%a mean He4 lattice spacing $a_0=3.6\AA.$
From these, we find $\kappa b^2 =3.7\times 10^2 K \AA.$ 

We have used\cite{norbert} $\theta=0.2$ degree.
The linear spacing between the dislocations is thus 
$a_x=a_0/\theta = 1029 \AA.$
From an estimate of the dislocation density of $6\times 10^9/cm^2=1/(a_xd),$  
we estimate a mean spacing between the boundaries as $d=10^4\AA/6.$
The core energy is not known.
In two dimensional melting,
it is suggested\cite{chuimelt} that when $E_c >E_{c0}=0.056 \kappa a_0,$
the transition becomes first order. This provides a sense of scale
for $E_c.$ In fig. 2 we show the roughening temperature for the x
and the y modes (dashed and solid lines respectively)
as a function of $E_c$. When $E_c$ is around $E_{c0},$
our estimate of $T_R$ for the x mode
is of the same order of magnitude as the experimental
transition temperatures of Day and coworkers\cite{jb}. $T_R$ 
for the y mode is of the same order as that of Burns and
coworkers\cite{x} and not a strong function of $E_c$.
The experimental results are indicated by the dotted lines. 
We next examine the physical implications of the transitions. 

We first show that only the upper y transition is connected with
the large scale orientation fluctutaion of the transition.
Our calculations is connected with the fluctuation of the position
of the dislocations. We first relate these to the fluctuation
of the atomic positions.
The displacement $\bu$ of the crystal
at $\br$ caused by a dislocation located at position 
{\bf c} can be written in Fourier transform as
${\bf u(r-c)}=\int d{\bf k} \exp[i{\bf k\cdot (r-c)}] {\bf u}_k$ 
where
$${\bf u}(k)=b[ {\bf e}_y/k^2-0.5(1-2\sigma)/(1-\sigma) {\bf k}k_y/k^4.]$$

$\sigma$ is the Poisson ratio.
The change in the displacement as the dislocations on the two boundaries
are moved from $\bc$ to $\bc+\delta \bc$ contains contributions from both
grain boundaries and is
equal to $\delta \bu =\delta \bu_1(\br)-\delta \bu_2(\br);$
$\delta \bu_1(\br)=\sum_j [\bu(\br-\bc_j-\delta \bc_j)-\bu(\br-\bc_j)],$
$\delta \bu_2(\br)=\sum_j [\bu(\br-\bc_j-{\bf e}_y d-\delta \bc_j)
-\bu(\br-\bc_j-{\bf e}_y d)].$ $\bc_j$ is in the xy plane.

Let us look at the angular fluctuation.
We have $\delta \theta=(-\partial_x \delta u_y+\partial_y \delta u_x)/2.$
For the "parallel" wave vector $k_p$ in the xz plan,
we are interested in $\int dy <(\delta\theta)_{k_p} (\delta\theta)_{-k_p}> 
=\int dk_y<|\delta \theta_{k_p,k_y}|^2>.$
%For the y direction,
%there may not be a periodic lattice. $\delta c_{ky}
%=\delta c(y=0)+\exp(-ik_y d)\delta c(y=d).$ 
%We are interested in the acoustic mode with $\delta c(y=0)\approx \delta c(y=d).$
%Thus $\delta c_{ky} \approx 2 \delta c(y=0)$ for small $k_y$.
It is straightforward to show that $\delta \bu(k)= {\bf u}(k) [g(k)-g'(k)]$
where $g-g'\approx k_y d \sum_G {\bf k\cdot \delta c_{k_p}} .$
We finally obtain, with a = x, y,
\begin{equation}
\int dy<(\delta\theta_{k_p})^2>=
\sum_{a,G}F_a(k) <|\delta c_a(-k_p-G)\delta c_a(k_p+G)|>.
\end{equation}
$F_a(k)=\int dk_y [-k_x u_y(k)+k_yu_x(k)]^2(k_yd)^{2}k_a^2,$
$G$ is a reciprocal lattice vector.
Thus the angular fluctuation of a grain
can be related to fluctuation of the position of the GB.
%We have two phonon modes, with a=b=y or a=b=x.
%So we have terms of the form
%$$\int d^2 k'_p F_a /(A_yk'_x^2+B_yk'_z^2)$$
%$k'=k+G.$
Subtituting in the expression for $\bu (k),$
we get $$F_a=b^2k_x^2\int dk_y (k_yd)^{2}k_a^2 /k^4.$$
%Note that $k^2=k_y^2+k_p^2,$ $k_p^2=k_x^2+k_z^2$ with a nonzero $k_y$!

%We expect $\int d k_yk_y^{2+x}/k^4
%\approx \delta(x)O(1/k_p)+\delta(x-2)O(\pi/a) ).$
Thus, for $k_z=0,$
\beq
\int dy<(\delta\theta_{k_p})^2>\approx
 k_x^3 <|\delta c_x(-k_x)\delta c_x(k_x)|>
%/(A_yk_x^2+\Delta))
\eeq
 for the x mode
\beq
\int dy<(\delta\theta_{k_p})^2>\approx
 k_x^2\pi <|\delta c_y(-k_x)\delta c_y(k_x)|>/a % /(A_yk_x^2+\Delta)
\eeq
for the y mode. 
 
From previous calculation of the roughening transition\cite{chui},
in a purely relaxational model, 
$$
<|\delta c(-k_x)\delta c(k_x)|>
\approx 1/[A(k_x^2+\xi^{-2})].
$$
As the transition is approached, the "elastic" coefficient $A$ 
and the relaxational rate are expected
to exhibit square root cusps; with $\xi\approx \exp[c/(T_R-T)^{1/2}]$
for $T<T_R$.

In the roughened phase with $\xi^{-1}=0,$ 
$<|\delta c(-k_x)\delta c(k_x)|>~1/(Ak_x^2).$ As $k_x$ approaches zero,
$\int dy<(\delta\theta_{k_p})^2>$ becomes finite above the higher temperature
roughened phase but remains zero between the two roughening transitions.
This is consistent with our interpertation that the higher roughening 
transition for the y mode corresponds to the observation from recent
X-ray measurements\cite{x}
Above the x mode roughening transition, 
because of the change in mobility of the dislocation,
a change in the elastic coefficients results; similar to that 
observed experimentally\cite{jb}.

Experimentally the addition of $^3$He in ppm to ppb concentration changes
the temperature dependence of the elastic modulus anomaly.
The core energy can be changed. As we see in fig. 2, 
$T_R$ can be a sensitive function of $E_c$.
The impurities can also change the effective magnitude of the pinning potential
as follows. We represent the pinning potential between $^3$He atoms at
${\bf s}_a$ and the dislocation line elements by 
$V^i=\sum_a W(\bc_j-{\bf s}_a).$
This appears as a factor $[\exp(-V^i/kT)]_{av}$ in the partition function
where the square bracket means an impurity averaging over the positions
${\bf s}_a$. If we approximate this factor by an cumulant expansion, we 
get to lowest order a term of the form $\exp([V^i]_{av}/kT).$ Because
the $^3$He appears as a substitional impurity, $[V^i]_{av}$ has the same
periodicity as the lattice and just modifies the effective strength
of the Peierl's potential $U_{x,y}.$  The next order in the cumulant expansion
provides for an effective short range attraction between the dislocation
lines, which can modify the elastic constants $C$ and hence $T_R.$
This short range interaction
is probably weaker than the long range elastic interaction in eq. (1).

The main difference between the current system and the pinning of flux line
lattices (FLL) in superconductors
and incommensurate charge density wave (CDW)
systems is the presence
of the commensurate pinning potential in the current system so that the
"ground state" is simple. For the CDW and the FLL system, the focus 
is on the nature of the ground state as a result of the competition
between the random pinning potential and the elastic energy cost to 
distort the lattice. In those systems, the interest is usually on three
dimensional objects whereas the current system is two dimensional.

%The movement of dislocations can induce ring-like exchanges.\cite{book}
%The dislocation motion may also form waves.\cite{d_wave} However, these
%phenomena cannot occur when they are pinned and a "gap" exists 
%in the excitation spectrum.
 
In conclusion we found that grain boundaries can roughen; there are
two transitions. Estimate of their physical properties and their transition 
tempertaures are consistent with recent experimental observations.
We do not completely understand what is the connection between
grain boundary roughening and supersolid behavior. We can think of many 
scenarios but further work is necessary to clarify if any of them is valid.

STC thanks N. Mulders for helpful discussions.

%For $\Delta=0$ and $k_x=0,$ 
%this is divergent for the y mode but not for
%the x mode! For the y mode, as the transition is approached
%we expect, for $k_z=0$,
%$$\int d y<(\delta\theta_{k_p})^2>\approx k_x/\Delta$$ for $T<T_R$ where 
%$\Delta$ is the gap. For $T>T_R,$
%$$\int dy <(\delta\theta_{k_p})^2>\approx \pi/(a A).$$ 

%Assuming that $c_j$ is small, we get
%$f(k)\approx \sum_j [1-i{\bf k\cdot c_j}].$ 
%Now ${\bf c}_j={\bf e}_{x,y}\exp(ip\cdot R_j)$ for the
%two polarizations.

%%% ----------------------------------
%%\begin{figure}
%%%\includegraphics[height=12cm,width=0.8\textwidth]{fig1}
%%\includegraphics{fig1}
%%\caption{\label{fig1}
%%Geometry of the scattering problem}
%%\end{figure}
%%% ----------------------------------


\begin{thebibliography}{99}
\bibitem{chan}
E. Kim and M. W. H. Chan, Science 305 1941 (2004). 
\bibitem{x}
C. A. Burns, N. Mulders, L. Lurio, M. H. W. Chan, A. Said, C. N. Kodituwakku
and P. M. Platzman, Phys. Rev. B78, 224305 (2008).
\bibitem{jb}
J. Day and J. Beamish, Nature (London) 450, 853 (2007).
\bibitem{balibar}
S. Sasaki, R. Ishiguro, F. Caupin, H. J. Maris, and S. Balibar, 
Science 313, 1098 (2006).
\bibitem{reppy}
A. S. C. Rittner and J. Reppy, Phys. Rev. Lett. 98, 175302 (2007).
\bibitem{dis}
M. Boninsegni, A. B. Kuklov, L. Pollet, N. V. Prokofev, 
B. V. Svistunov and M. Troyer, Phys. Rev. Lett. 99, 035301 (2007).
\bibitem{toner}
J. Toner, Phys. Rev. Lett. 100, 035302 (2008).
\bibitem{lagb}
L. Pollet, M. Boninsegni, A. B. Kuklov, N. V. Prokofev, B. V. Svistunov and M. Troyer, Phys. Rev. Lett. 98, 135301 (2007).
\bibitem{BM} L. Bonsall and A. Maraduddin, Phys. Rev. B
15, 1959 (1977).
\bibitem{chui} 
S T. Chui and J. D. Weeks, Phys. Rev. Lett. 40, 733 (1978).
\bibitem{elas}
L. Nosanow and N. R. Werthamer Phys. Rev. Lett. 15, 618 (1965);
L. Goldstein, Phys. Rev. 128, 1520 (1965).
\bibitem{norbert}
N. Mulders, private communication.
\bibitem{chuimelt}
S. T. Chui, Phys. Rev. B28, 178 (1983).
%\bibitem{book}
%"Physics of the electron solid", Fig.3.8, p. 30, ed. S. T. Chui, 1994, 
%International Press, Boston.
%\bibitem{d_wave}
%K Esfarjani,
%S. T. Chui and X. Qiu, Phys. Rev. B46, 4638 (1992).
\end{thebibliography}
\end{document}